\documentclass[journal]{IEEEtran}
\usepackage{cite}
\usepackage{amsmath,amssymb,amsfonts}
\usepackage{algorithmic}
\usepackage{graphicx}
\usepackage{textcomp}
\usepackage{soul}
\usepackage{bm}

\makeatletter
\AtBeginDocument{\DeclareMathVersion{bold}
\SetSymbolFont{operators}{bold}{T1}{times}{b}{n}
\SetMathAlphabet{\mathrm}{bold}{T1}{times}{b}{n}
\SetMathAlphabet{\mathit}{bold}{T1}{times}{b}{it}
\SetMathAlphabet{\mathbf}{bold}{T1}{times}{b}{n}
\SetMathAlphabet{\mathtt}{bold}{OT1}{pcr}{b}{n}
\SetSymbolFont{symbols}{bold}{OMS}{cmsy}{b}{n}
\renewcommand\boldmath{\@nomath\boldmath\mathversion{bold}}}
\makeatother

\def\BibTeX{{\rm B\kern-.05em{\sc i\kern-.025em b}\kern-.08em
    T\kern-.1667em\lower.7ex\hbox{E}\kern-.125emX}}

\begin{document}

\title{Internet of Paint (IoP): Design, Challenges, Applications and Future Directions}


\author{Lasantha~Thakshila~Wedage,~\IEEEmembership{Graduate Student Member,~IEEE}, Mehmet C. Vuran,~\IEEEmembership{Member,~IEEE}, Bernard~Butler,~\IEEEmembership{Senior Member, IEEE}\thanks{\it{Lasantha Thakshila Wedage and Bernard Butler are with the Walton Institute, South East Technological University, Ireland.}},\thanks{\it{Mehmet C. Vuran and Sasitharan Balasubramaniam are with the University of Nebraska-Lincoln, USA.}} Christos~Argyropoulos,~\IEEEmembership{Senior Member,~IEEE} \thanks{\it{Christos Argyropoulos is with the Pennsylvania State University, USA.}} and Sasitharan~Balasubramaniam,~\IEEEmembership{Senior Member,~IEEE}} 


\maketitle



\begin{abstract}
The proliferation of nano-technology has enabled novel applications in various fields, including the potential for miniaturized terahertz (THz)-enabled devices. The extra-large bandwidth available in the THz spectrum can facilitate high-speed communication, even for transmission through lossy media. Culminating these capabilities, this paper introduces a new paradigm: \emph{Internet of Paint} (IoP), transforming the 40,000-year-old concept of paint with extensive connectivity and sensing capabilities. IoP will enable seamless, massively parallel, and high-capacity communication and sensing capabilities, enabling innovative applications. IoP is expected to offer communication and sensing using nano-devices embedded in paint. In this paper, challenges towards the realization of IoP are discussed. Simulations indicate that THz signals in this stratified medium can form lateral waves that propagate along the Air-Paint interface and that relatively high channel capacity can be achieved by optimizing the transceiver locations, the paint thickness, and color. Addressing the challenges regarding nano-devices, nano-transceivers, materials, antennas, and power for IoP holds the potential to transform communication technologies and their seamless integration with living spaces. 
\end{abstract}



\section{Introduction}
\label{Intro}

Nanotechnology was initially pitched by Nobel laureate physicist Richard Feynman in his speech ``\emph{There’s plenty of room at the bottom}" at the annual meeting of the American Physical Society (APS) in $1959$ \cite{feynman1959there}. Feynman laid out a vision in which materials and components could be engineered and manufactured at the nanoscale. While the last few decades has seen the field of nanotechnology focus on novel materials and components, we are now starting to witness the emergence of novel \emph{nanodevices} that can be assembled from these components. This has led to the field of nano communications, where the nanodevices that have only limited capabilities can communicate and network with each other \cite{IoNT_Akyildiz}. The Internet of Nano Things (\emph{IoNT}) envisions these nanodevices connecting to cyberspace, where these devices can be embedded into plants \cite{giraldo2019nanobiotechnology} and in human as well as animal tissues \cite{solaimuthu2020nano,li2021diagnosis,sunwoo2021wearable}  and collecting vast quantities of data at molecular scale. In this context, the nanodevices of the IoNT can serve as transceivers, sensors, or actuators. Building on these innovations, we introduce a new concept that takes IoNT to a new level: Internet of Paint (IoP), where nanodevices are embedded seamlessly into paint and are able to form communication networks that is part of the medium.

Paint has been used since pre-historical times. Nearly all man-made objects require a surface coating for protection (from moisture, UV rays, corrosion, etc.) and/or decorative purposes. We envision how commonly used paint coating technologies can be enhanced with additional purpose and capabilities if a communication infrastructure is integrated into the medium. Accordingly, objects such as walls, ceilings, furniture, and even consumer goods could be integrated into the Internet of Things (IoT), while maintaining their aesthetics and function. The IoP enables novel applications that leverage nanoscale communication and networking infrastructure deployed in the paint and provide a new mode of communication and sensing within the paint medium. The IoP offers huge potential for extending wireless coverage in underserved areas by covering surfaces with functional paint and for enabling new applications (see Fig. \ref{fig:Main_figure}) by combining nanodevice-enabled communication and sensing.

In the IoP, the nanodevices are suspended in the paint mixture before it is applied to a surface. Once the mixture is applied onto a surface, the nanodevice embedded within the paint medium can communicate with each other. The paint layer still offers protection and decoration to the objects it covers, but the embedded nanodevices add communication capabilities that were not there before. This paper considers the challenges and opportunities offered by using the most common paint coating as the medium in which the nanodevices are embedded. These challenges include (i) ensuring that the dimensions of these devices are seamlessly integrated into the paint so it doesn't change the form factor of the applied paint, and (ii) determining the communication properties of the nanodevices with antennas at nanoscale. Since the nanodevices function as wireless transceivers and are embedded inside the paint medium, the dimensions of the antenna needs to be significantly smaller than the thickness of that medium. This is to avoid disrupting the look and normal function of the paint layer. Therefore, the relationship between the corresponding wavelength and the antenna dimensions needs to be considered when selecting the frequency range for communication within the paint medium. Indeed, frequencies from sub-Terahertz (sub-THz) as well as Terahertz (THz) frequencies meet the dimensions of the required antenna and at this miniature size can be embedded into paint layer in an unobtrusive manner.
However, a number of media (e.g., paint and other building materials) as well as certain atmospheric gas compositions tend to absorb THz waves. The channel losses in these frequencies are known to be high, which results in very short distance communication between the nanodevices.

Based on the challenges described above, as well as considerations in components and materials that will be used to construct nanodevices for IoP, a number of factors will also need to be considered. The components and materials used for the antenna of the nanodevices will be embedded into various paint types that may be used to serve a specific purpose. For example, certain paint mixture can have specific composition to improve radiative cooling on the walls. The mixture can have an impact on the radiative behaviour of the signals in the sub-THz and THz spectrum, which can also impact on the materials that are selected for the antenna to used in the IoP. A candidate antenna material that has been proposed for IoT nanodevices, and communicates in the sub-THz and THz frequencies, are metamaterials that can be made of emerging ultra-thin materials such as graphene. Another factor that has to be taken into consideration is the substrate and its dielectric properties of that material. The paint medium will provide a coating on the antenna, which can function as a dielectric layer (e.g., superstrate) above and below it. When properly designed, this layer of paint materials on the antenna can enhance the antenna performance as well as provide  protection \cite{khan2020high}. THz communication within and through the paint has recently been explored~\cite{Thakshila_IoP_2024}, revealing unique characteristics in this type of medium. IoP-enabled wireless communication has the potential to enable unique applications that have not been possible at this scale before. 

The rest of the paper is organized as follows: In Section~\ref{sec:IoP_and_Applications}, potential IoP applications in the context of indoor, outdoor, and space environments are discussed. The device characteristics needed to realise these opportunities are described in Section~\ref{sec:Device_Model}. In Section~\ref{sec:Characteristics_of_THz_Comms_in_IoP}, early work in addressing IoP communication challenges are discussed. Diverse challenges for the realization of IoP are discussed in Section~\ref{sec:IoP_Challenges}. The paper concludes in Section~\ref{sec:Conclusion}.

\section{IoP Applications}
\label{sec:IoP_and_Applications}

\begin{figure*}[t!]
    \centering
    \includegraphics[width=\linewidth]{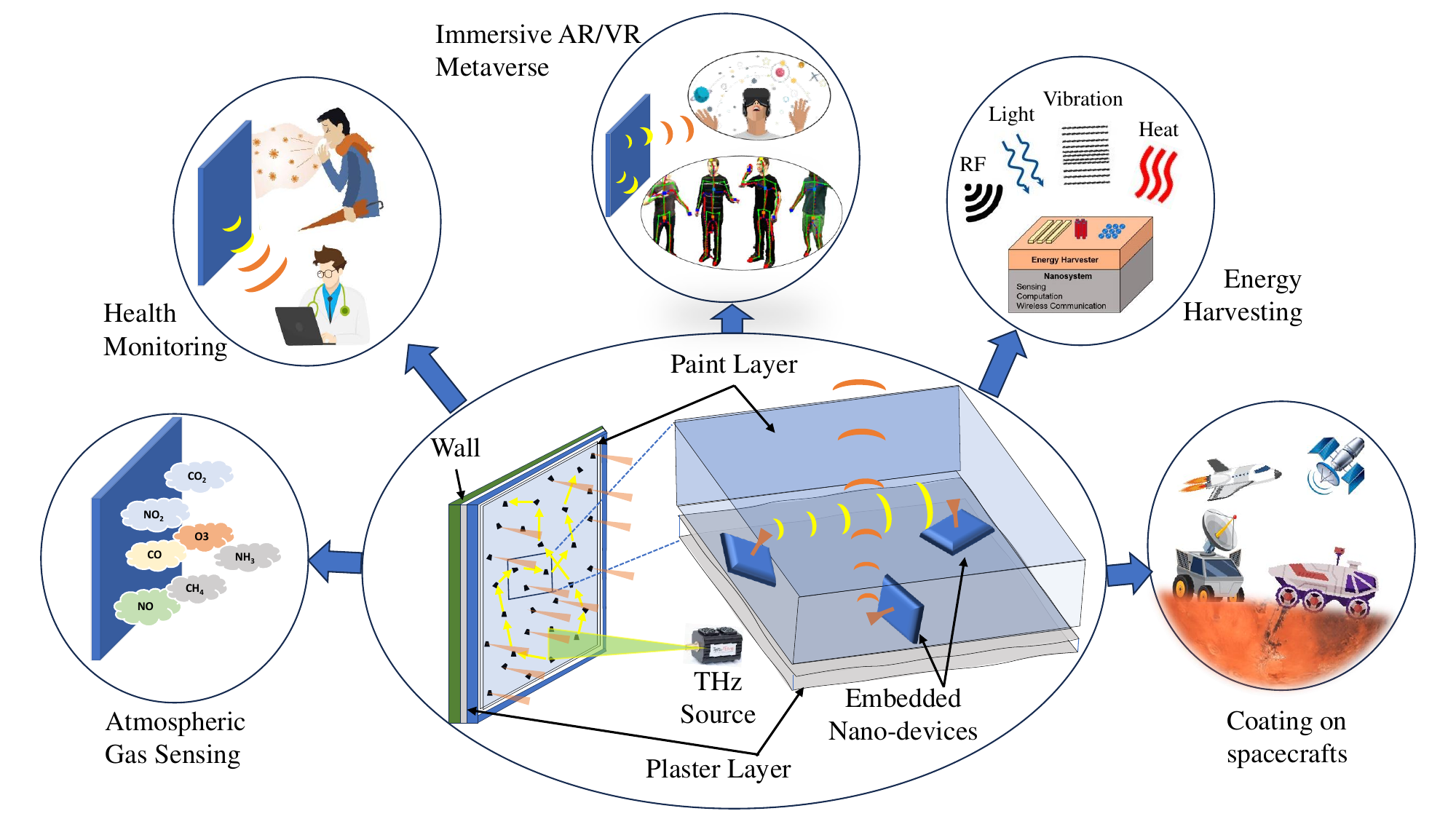}
    \caption{Internet of Paint (IoP) and its futuristic applications.}
    \label{fig:Main_figure}
\end{figure*}

Potential IoP applications are illustrated in Fig.~\ref{fig:Main_figure}. In this section, we discuss applications that would transform paint and coatings into communication and sensing hubs, transcending the traditional separation between form and function. The applications offer a future where painted surfaces enhance communication and sensing indoors and outdoors, as well as in space-based applications.

\subsection{Indoor Applications}
\label{subsec:Indoor}

    \textbf{Wall communication infrastructure}: The IoP offers a new paint-based communication infrastructure in the form of a reflecting surface or a hub, where nano-devices are used to reflect and steer THz signals. IoP devices could be constructed using nanomaterials that enable devices to be non-intrusively integrated into the paint medium and appear invisible. Accordingly, bulky elements used in communication devices or IRS could be minimized, reducing deployment costs. Further, nano-networks integrated into paint can allow signals to travel along the wall. This will require new routing mechanisms that suit the energy limitations of the devices, and that take account of how THz signals propagate through paint.
    
    \textbf{IoP for Public Health Monitoring}: The COVID-$19$ pandemic increased the demand for pathogen-sensing technologies. The IoP could facilitate the sensing of viral particles expelled when coughing or sneezing. THz signals might be used to sense viral particles adhering to the wall. This could lead to large-scale data collection that tracks the movement of virus particles, especially in medical centres and in public areas where people congregate (e.g., on buses, trains, etc.).  

    \textbf{Posture Recognition}: The sub-THz and THz signal could be used for sensing purposes (i.e., RF-based sensing), where the relatively short wavelengths significantly improve spatial accuracy. Such refined spatial sensing offers opportunities for gaming, posture recognition, head recognition and headset sensing for AR/VR, position sensing and immersive media for the metaverse and its avatars.

    \textbf{Back-scatter Monitoring}: The nanodevices in the paint can be powered using external sources (e.g., ultrasound or EM waves). External sources could emit waves to power the devices and sense back-scattered waves to assess the location and condition of the embedded nanodevices.

\subsection{Outdoor Applications}
\label{subsec:outdoor}

The indoor applications above might also be relevant in outdoor settings. Additionally, IoP could enable the following applications.

    \textbf{IoP for Airborne Gas and Chemical Detection}: The nanodevices embedded into the paint mixture could be used to collect data using point-to-point THz signalling to sense gases \cite{Wedage_2023}. This might be used to sense and track long-term greenhouse gas changes over large geographical areas, for climate change. Moreover, THz short pulses could be utilized to detect bacterial spores and chemical materials, opening the door to powerful and flexible sensors.

    \textbf{Efficient Energy Harvesting}: Nanodevices embedded in the paint will need to harness ambient energy from sources, such as light or vibration, to replenish the energy of the embedded devices, thereby enabling sustainable communication infrastructure. Moreover, energy harvesting would reduce the reliance on external power sources, reducing the IoP's impact on the environment.

\subsection{Space Applications}
\label{subsec:space}

\textbf{Coating on Spacecraft}: IoP could enable high data rate pervasive communication \emph{within} a spacecraft. This will motivate new research directions exploring advanced coating technologies for enhanced performance, protection and communication on space vehicles. Additionally, the IoP could be used to detect surface abnormalities or cracks on spacecraft, alerting the crew to structural damage during space missions. Finally, nanoparticles in paint designed to provide radiative cooling is the most promising passive way to regulate the temperature of satellites and other spacecraft \cite{Chris2022mechanically}.

\textbf{IoP for Extraterrestrial Gas and Chemical Detection}: Paint layers incorporating nanodevices could be employed on planetary surface monitoring vehicles, such as Mars Rovers. Such vehicles would perform atmospheric gas sensing with limited additional sensing infrastructure, and would be able to share this information quickly for later analysis. 

\section{Nano-transceivers for IoP}
\label{sec:Device_Model}

Nanotechnology adoption accelerated a few decades ago with the discovery of innovative materials and component designs. This resulted from enhanced chemical and physical materials, as well as integration into miniature devices. In this section we discuss the nanotechnologies that have transformed paint by making it more robust and durable in the presence of variable environmental conditions. This is followed by a brief discussion on proposed nano-transceivers for THz communication, considering the materials used for designing antennas as well as their applications. Lastly, we discuss how the development of nanotechnologies in paint and nano-transceivers for THz communications might help to design future IoP transceivers. 

\begin{table*}[t!]
    \centering
    \caption{Properties of polymers that are commonly used as flexible substrates for THz metasurfaces.}
    \label{tab:Properties_of_polymers}
    \begin{tabular}{|l|c|c|c|c|c|c|}
    \hline
    Material & Dielectric  & Loss & Absorption  & Refractive & Frequency & Reference \\
      &constant & tangent & coefficient (cm$^{-1}$) & index & range (THz) & \\
    \hline
     Polyethylene terephthalate (PET)   &2.86 & 0.053-0.072& 25& &0.2-2.5 & \cite{walia2015flexible}\\
     Polyethylene naphthalate (PEN)   &2.56 &0.003 &1 & &0.2-2.5 & \cite{walia2015flexible}\\
     Polymethyl methacrylate (PMMA)   &2.22 & 0.042-0.07 & 22&1.49 & 0.2-2.5& \cite{walia2015flexible} \\
     Polypropylene &3 & 0.12&2 & &0.2-2.5 & \cite{walia2015flexible}\\
     Polytetrafluoroethylene (Teflon) & 2.39 &   & 1.6 & 1.42 & 1 & \cite{Teflon2012studying}\\
      \hline    
    \end{tabular}
    
\end{table*}

\subsection{Nanotechnology in Paint}
\label{subsec:Nanotechnology_in_Paint}

Conventional paint is developed by mixing colour pigments ($20\%$-$30\%$) in a liquid medium. Additional solvents ($30\%$-$50\%$) and bulking agents ($20\%$-$50\%$) are added to achieve the required functional properties of the paint to suit environmental conditions. Such properties might include resistance to UV damage, heat, or damage from chemicals or abrasion from other surfaces. Therefore, the THz interaction of dried/cured paint films is influenced by the collective behaviour of the pigments (e.g., type and size $\in [0.05\,\mu m, 1\,\mu m]$), the binders, and other additives. 

Recently, nanomaterials have been used to enhance the properties of paint. For example, scientists have developed lightweight paint ($0.4$\,g/m$^2$) using micro-scale ($100$\,nm diameter) aluminum pigment particles produced with an ultrahigh vacuum electron beam evaporator, with sub-wavelength plasmonic cavities mixed into the composition to guarantee a high degree of angle insensitivity of paint colour~\cite{cencillo2023ultralight}. The plasmonic nanocavities are made of metals, such as aluminium, and can scatter light signals at different wavelengths leading to a variety of colours from the paint pigment. This is also called structural colour paint. Moreover, a new paint type has been introduced in \cite{yedra2016conductive} with electrical conductive properties. This paint has a polymer matrix in which multiwall carbon nanotubes (MWCNTs) are integrated, which can absorb UV light, thereby minimizing photodegradation of the resin. MWCNTs have complex dielectric properties that can influence THz wave propagation depending on their diameter, length and alignment. In addition to their intrinsic conductivity, MWCNTs can exhibit strong dielectric response in the THz frequency range due to interband transitions and electronic confinement effects. Because of their conductivity and resonance effects, MWCNTs can also absorb and scatter THz radiation. Nanocoating is an emerging development in the paint industry that leverages the unique properties of nanomaterials to create paints that improve durability, enhance weather resistance (UV radiation, temperature fluctuations, etc.), and improve adhesion. 

The nanotechnologies used to transform paint materials can be exploited when designing IoP transceivers. In particular, we can integrate nanomaterials with a low refractive index for THz signals into the devices. Furthermore, materials such as carbon nanotubes could be used to construct IoP transceivers because they have lower absorption loss than the surrounding medium. 

\subsection{Nano-transceivers for THz Communications}
\label{subsec:Current_Devices}

Nanocommunications has led to novel nano-transceiver devices that integrate nanomaterials and components, such as nanoantennas produced from metamaterials \cite{ChrisrefId0_2015}. We briefly discuss some designs that could lay a foundation for future IoP transceivers.

Graphene is a $2$D conductor metamaterial that has been used to produce plasmonic antennas that communicate in the THz band. One example is the Graphene-enabled Wireless Networks on-Chip communication within computer processors that enables cores to communicates wirelessly using THz signals~\cite{graphene2013}. Plasmonic nanoantennas for nano-transceivers have been proposed for integrated communication as well as bio-sensing for intra-body communication for medical applications~\cite{ChrisrefId0_2015}. However, transceivers for IoP need to integrate materials compatible with THz communication and the surrounding paint. Requirements include materials suited to antenna design, a paint mix that does not cause excessive THz attenuation, packaging that is compatible with the paint and provides protection for the IoP transceivers inside. We discuss this in the following subsection. 

\subsection{Materials for Paint and Transceivers}
\label{subsec:Novel_Devices_for_IoP}

For nano-transceivers that communicate at THz frequencies, several design changes need to be considered for IoP. New materials suitable for paint and capable of emitting THz signals will be needed, as well as reflecting and refracting along the wall. In Table \ref{tab:Properties_of_polymers}, various polymers are shown with their dielectric constants, loss tangents, absorption coefficients, refractive indices, and their corresponding THz frequency ranges. 
The table shows the frequency ranges that can operate in these polymers, indicating that these frequencies could be used for communication. A future design objective will be to choose paint formulations that have (a) low THz loss and (b) are either protective or chemically compatible with the embedded IoP nano-transceivers. For example, it will be necessary to cover the nano-transceivers with materials discussed in Section~\ref{subsec:Nanotechnology_in_Paint} (e.g., MWCNTs) that can contribute to enhancing the paint material, while minimizing THz attenuation. Non-volatile water-based paints can also be blended with polymers such as polyethylene and polypropylene to provide bulk and aid adhesion \cite{nakayama1998polymer} to facilitate a stable substrate for embedded IoP nano-transceivers. Combining the THz antennas with paint-based nanomaterials could lead to low permittivity and loss tangent properties. These materials could also affect the directional refraction and reflection of THz signals in the paint medium (We evaluate the performance of these waves in Section~\ref{sec:Characteristics_of_THz_Comms_in_IoP}). They can also be used to change the input impedance of the resulted nano antennas, causing resonant shifts, i.e., operation at different wavelengths.

\section{Characteristics of THz Communications in IoP}
\label{sec:Characteristics_of_THz_Comms_in_IoP}

\begin{figure}[t]
    \centering
    \includegraphics[width=8cm]{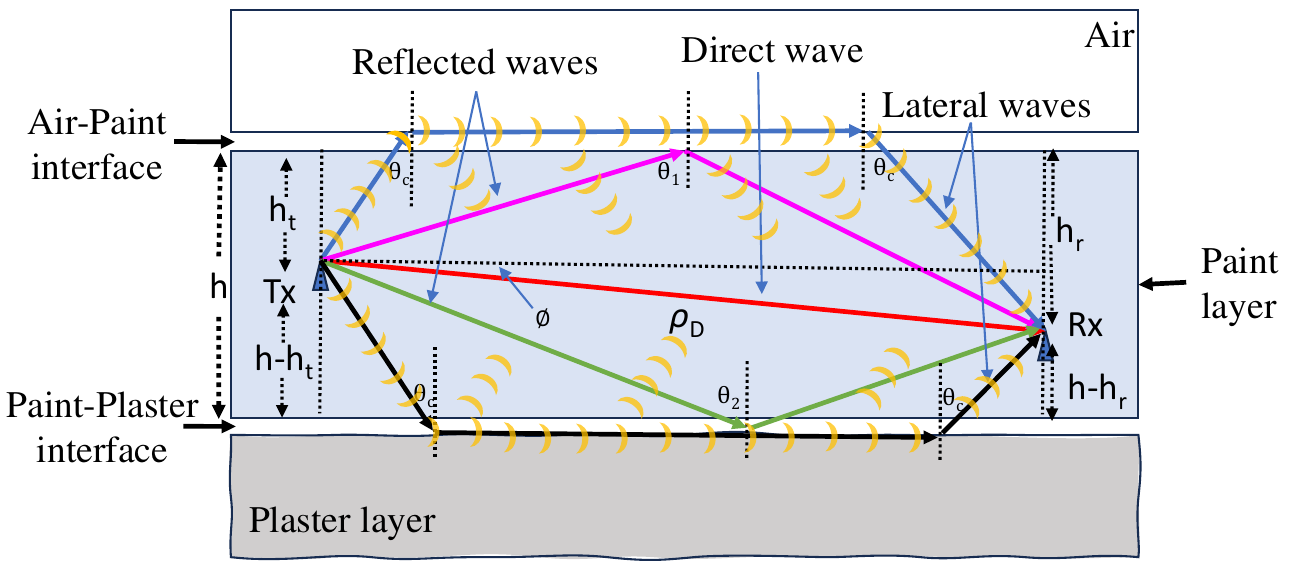}
    \caption{Illustration of device-to-device communication for nano-devices embedded in paint resulting in the direct path, reflected paths, and lateral waves that can propagate along the air-path and paint-plaster interfaces~\cite{Thakshila_IoP_2024}.}
    \label{fig:Channel_model}
\end{figure}

THz channel modelling in paint is a new direction in the field of IoNT. In~\cite{Thakshila_IoP_2024} we developed a point-to-point THz channel model between THz-enabled nanodevices embedded in paint over the plaster surface of a wall. Nano-devices embedded in the paint layer produce electromagnetic (EM) wave propagation paths for three different media: air, paint, and plaster. The speed of EM waves differs in media having different medium properties. For instance, EM wave speed in paint ($c_p$) is reduced by the factor of the refractive index ($n_p$) relative to the vacuum ($c$) (i.e. $c_p \equiv c/n_p$). Furthermore, the three different layers create two interfaces, which are Air-Paint (A-P) and Paint-Plaster (P-P). An EM wave that is incident on each of these interfaces undergoes reflection and refraction, resulting in a grouped wave phenomenon called \emph{lateral waves}~\cite{vuran2018internet}, according to Snell's law.

Thus, IoP has the potential for multipath communication between nano-devices in the paint layer. Indeed, there are five canonical paths because of the higher refractive index of paint compared to air and plaster. The five dominant waves are: (i) direct wave (DW), (ii) reflected wave from the Air-Paint interface (RW-A), (iii) reflected wave from the Paint-Plaster interface (RW-P), (iv) lateral wave along the A-P interface (LW-A), and (v) lateral wave along the P-P interface (LW-P). LW-A and LW-P are also known as head waves since they are the first waves to reach the receiver. When a wave reaches the A-P interface at a particular angle known as the critical angle according to Snell's law, it changes the wave direction and propagates through the air but closer to the upper paint layer. It leaks the EM waves' energy back into the paint layer (see Fig. \ref{fig:Channel_model}) in the critical angle direction while transmitting through the interface until it reaches the receiver. 

\begin{figure}[t]
    \centering
    \includegraphics[width=7cm]{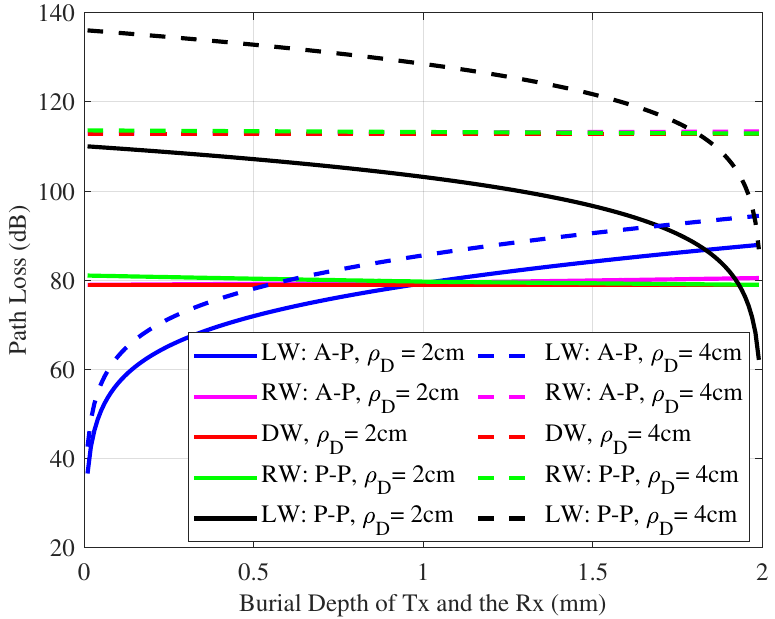}
    \caption{Path loss variation with the burial depth for different LoS distances ($\rho_D=2$ and $4$\,cm) for a fixed frequency ($200$\,GHz) and thickness ($2$\,mm) for Titanium white paint (n = $2.13$) (LW: Lateral Wave, RW: Reflected Wave, DW: Direct Wave, A-P: Air-Paint and P-P: Paint-Plaster Interface)~\cite{Thakshila_IoP_2024}.}
    \label{fig:pathloss_vs_LoS_dis}
\end{figure}

The path loss corresponding to each dominant wave is due to spreading and medium absorption losses. The DW, RW-A and RW-P propagate entirely through the paint medium. However, the roughness of the paint and plaster surfaces should be taken into account for the RW from both the A-P and P-P interfaces. Moreover, according to Snell's Law, total internal reflection results in only one of the multiple RWs reaching the receiver, which is incident upon the interfaces exceeding the critical angle. The LW-A path propagates through both paint and air. Thus, spreading in both mediums (A-P), absorption by paint, and absorption by atmospheric gases should be considered when evaluating the path loss. Note that the second medium for the LW-P path is plaster instead of air.

The burial depth (relative to the A-P interface) in the paint of each transceiver cannot be controlled, so it varies randomly. Thus, we performed an analysis of path loss, considering the direct, reflected, and lateral waves between transmitters and receivers buried at unequal depths in the paint layer~\cite{Thakshila_IoP_2024}. However, in this section, we present the simulation results for the scenario where the burial depths of the transceivers are equal (i.e., $h_t=h_r$, hence $\phi = 0$). This assumes that the necessary technology is available. Also, it i easier to discuss the simulations if \emph{both} transceivers are buried at the same depth, rather than at two separate depths. Figure \ref{fig:pathloss_vs_LoS_dis} shows how simulated path loss values vary with burial depth for different LoS distances ($2$ and $4$\,cm). In the Matlab\texttrademark{} simulations, the operation frequency of $200$\,GHz and thickness of $2$\,mm are used where the paint includes Titanium white pigments (n = $2.13$). A paint thickness of 2\,mm is chosen, which is thicker than most paint films, to represent worst-case scenarios. We notice that LW-A has the slowest rate of change of path loss, as the path loss of all the primary communication paths increases because spreading and absorption losses increase with distance. For instance, at a burial depth of $1$ mm, increasing the distance from $1$ cm to $4$ cm results in only a $12.79$ dB increase in LW-A path loss. In contrast, path loss increases of $53.89$ dB, $52.82$ dB, $52.39$ dB, and $41.97$ dB are observed for DW, RW-A, RW-P, and LW-P, respectively. Additionally, it is noted that LW-P only dominates the IoP channel when transceivers are positioned within $0.08$ mm of the P-P interface, while the LW-P path generally exhibits higher losses compared to the LW-A. This behavior can be attributed to the significantly higher absorption properties of plaster compared to the molecular absorption in air.

\begin{figure}[t]
    \centering 
    \includegraphics[width=7cm]{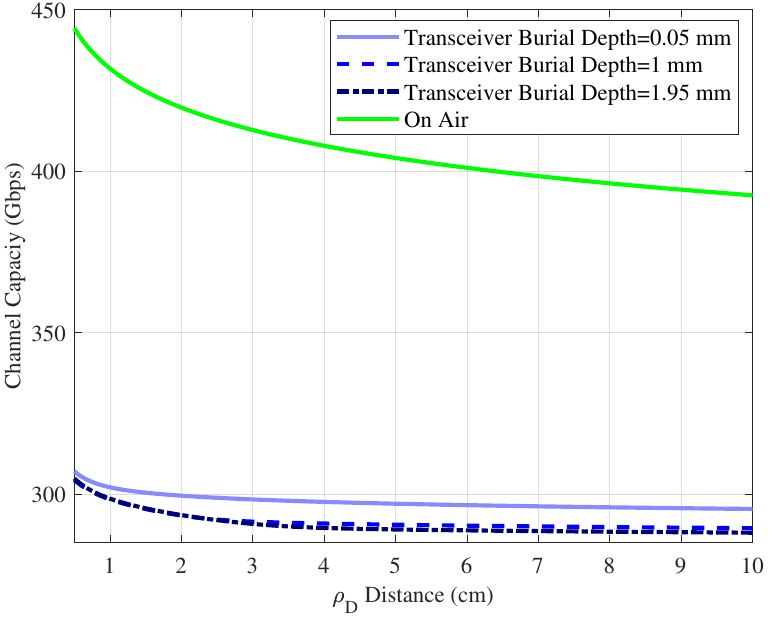}
    \caption{Comparison of Channel capacity through Air and Paint (Multipath) for various transceiver burial depths within $200$ to $300$\,GHz frequency. The thickness is considered as $2$\,mm for Titanium white paint (n = $2.13$) \cite{Thakshila_IoP_2024}.}
    \label{fig:CC_Air_paint}
\end{figure}

The channel capacity analysis can be used to predict performance metrics for the IoP. The total channel capacity of IoP channels at THz frequencies can be determined by dividing the total bandwidth into several narrow sub-bands and aggregating their individual channel capacities due to its strong frequency selectivity and the presence of non-white molecular noise~\cite{Thakshila_IoP_2024}. The simulated channel capacity values against LoS distances are illustrated in Fig. \ref{fig:CC_Air_paint} for the $200-300$\,GHz band while varying the burial depths of the transceivers $0.05$, $0.1$, and $1.95$\,mm, considering a $2$\,mm thickness paint (n = $2.13$) layer. When comparing channel capacity using IoP and air-based communication, the achievable channel capacity in paint is reduced by two orders of magnitude compared to air-based communication. This reduction can be reduced by careful topology design. When the transceivers are buried near the A-P interface, $1.9-15.8$\,Gbps higher channel capacity is observed compared to $0.1$, and $1.95$\,mm burial depths. The reason for the possibility of achieving relatively higher channel capacity is that most of the LW-A path passes through the lower refractive index medium of air. 

\section{IoP Challenges}
\label{sec:IoP_Challenges}

The IoP paradigm offers many advantages, but it also poses many challenges. Next, we discuss several challenges towards realizing IoP. 

\subsection{Medium Properties of Paint}
\label{subsec:Medium_Properties_of_Paint}

Similar to air-based communication, spreading and absorption losses are expected when communicating through paint. However, due partly to its relatively high density, paint has higher signal energy absorption compared to air. Although the spreading loss depends on frequency and distance between transceivers, it differs in paint compared to air due to a relatively high refractive index. This medium property causes the electromagnetic wave propagation speed to decrease in paint relative to air. Moreover, when establishing communication with the reflected wave from the A-P and P-P interfaces (see Fig. \ref{fig:Channel_model}), it is unrealistic to consider paint and plaster layers as smooth surfaces. Therefore, when developing communication channel models, it is essential to account for the roughness of both the paint and underlying plaster. Furthermore, when a suitable micro/nano patch antenna is embedded, the paint medium can function as a superstrate. Therefore, the dielectric properties of the paint need to be considered when calculating the resonance frequency and the bandwidth of the proposed antenna. Finally, the lifespan of the IoP network also relies on the properties of the paint. Damage to the paint layer limits the protection afforded to the embedded nanodevices, which can result in unreliable connectivity between the embedded nodes.

\subsection{Controlling the Placement and Orientation of Tranceivers}
\label{Subsec:Controlling_Placement_and_orientation}

 The burial depth and orientation (hence preferred signal propagation direction) of the nanodevices inside the paint layer is a significant engineering challenge. It is evident that if we could place the transceivers near the A-P interface, we could expect higher channel performance because the lateral waves have a higher Received Signal Strength than the corresponding waves that travel directly through the paint between the transmitter and receiver (see Section~\ref{sec:Characteristics_of_THz_Comms_in_IoP}). In contrast, if the pair of transceivers are buried at uneven burial depths, the signal travel a longer distance through the paint. Also, the orientation of the transceivers is significant because the THz band has higher free space path loss. Therefore, ``face-to-face'' directional antennas could increase the performance of the IoP.  

\subsection{Powering Nano-Devices}
\label{Powering_nano_devices}

Powering IoP nanodevices will be a major challenge. Since it is not practical to replace these nanodevices regularly, we expect that these IoP devices will be constructed from functionally stable nanomaterials, and will be able to harvest energy from their environment. An example is the use of nanocomponents that deform mechanically to harvest energy. One example is the use of zinc oxide nanowires that can be compressed by vibrations generated using an ultrasound source \cite{xu2010piezoelectric}. 

\subsection{Requiring a Dense Network}
\label{subsec:Require_Dense_network}

A dense network integrating many nano-devices is needed for IoP communication due to the above-mentioned orientation challenge, the varying thickness of the paint layer and the random, uneven spatial distribution of transceivers that were stirred into the paint rather than being placed in a regular pattern. When many devices are embedded in the paint, it creates a dense network with a high probability of having nearby, reasonably well oriented, transceivers for a connected network to form and hence to support reliable communication.   

\subsection{Impact of Noise}
\label{subsec:Molecular_absorption_and_Thermal_Noise}

Lateral wave propagation along the A-P interface (see Section \ref{sec:Characteristics_of_THz_Comms_in_IoP}) is the most promising communication path but is impacted by molecular absorption due to atmospheric gases. It is known that the internal vibrations of gas molecules emit EM radiation at particular frequencies, creating interference with the incident waves sharing those frequencies that induce this motion. Thus, molecular absorption introduces noise. Moreover, thermal noise generated by the nano-devices and home electrical appliances is also present in the communication environment. Therefore, when developing a channel model, these noise sources should be considered.

\subsection{Nano-device Response to Environmental Influences}
\label{subsec:Paint_impacted_by_the_Environment}

It is likely that nano-devices with graphene-based antennas will be susceptible to damage from moisture and chemical reactions. That being said, after being embedded in the paint, nano-devices are naturally protected from environmental moisture. To further safeguard the nano-devices, we propose to seal them with materials that are compatible with paint, such as the materials from Table \ref{tab:Properties_of_polymers}. Indeed, Teflon substrate, with a dielectric constant of 2.39 has high stability against temperature changes and high resistance to adhesion, can be used in to encapsulate graphene-based antennas to protect them. As specified in Table \ref{tab:Properties_of_polymers}, Teflon polymers can operate in certain THz frequency ranges. 

\subsection{Reflected Signal Coupling with Antennas}
\label{subsec:Reflected_Signal_Coupling_with_Antennas}

Simulation results (see Section~\ref{sec:Characteristics_of_THz_Comms_in_IoP}) suggest that placing the transceivers close to the A-P interface would help to achieve high performance through IoP. However, the placement of transceivers near the A-P interface is a challenge for communication because the reflected signals from the interface could start coupling with the antenna. This phenomenon has the potential to alter the current distribution of the antenna, changing its behavior. Potential detrimental effects include antenna impedance mismatch due to the coupling from the reflected waves. The impedance mismatch can affect the antenna radiation pattern and change its operational frequency. Ways to minimize this detrimental coupling effect exist by changing the antenna design or engineering its feeding circuits.

\subsection{THz Antennas in Paint}
\label{subsec:THz_Antennas_in_Paint}

Graphene exhibits conductive response at THz frequencies and is an ideal material to substitute bulky and lossy metals (e.g., copper) that are usually used in this type of antenna designs. As a result, graphene can be patterned to create ultrathin and flat patch antenna designs at THz frequencies~\cite{azizi2017terahertz}. However, the increased losses and design complexity of patterned graphene nanostructures is expected to be detrimental in the resulting antenna response. This is the main reason why only theoretical studies in this area exist to date \cite{azizi2017terahertz}. Optical nanoantennas are usually made of noble metals (gold, silver) patterned to create nanostructures (dipoles or patches) and offer an alternative to create THz antennas with extremely small profiles \cite{Chris8070946_2017}. However, the demanding nanofabrication requirements and high cost of producing such nanostructures make them prohibitive in commercial applications. Furthermore, feeding such nanoantenna designs is very challenging since signal generators in THz frequencies need further development. However, these challenges might be overcome in the future with new antenna engineering and design approaches as the research community addresses future 6G communications needs that include THz communication in paint.
    
\section{Conclusion}
\label{sec:Conclusion}

This article introduces the groundbreaking concept of the IoP, which enables high-speed communication through paint using embedded nano-devices. Despite lower data rates than air-based communication, IoP offers new opportunities for communication and sensing applications, taking advantage of new devices and multipath communication expected between embedded IoP devices. Some potential applications have been identified, opening new opportunities as devices are integrated into 
paint applied on everyday surfaces. Simulation results reveal that the lateral wave propagating through the A-P interface is the most promising communication path for IoP, and higher channel capacity can be achieved when the transceivers are placed near the A-P interface. We also highlight challenges that need to be overcome to establish reliable communication through the IoP. 

\section*{Acknowledgment}

This publication came from research conducted with the financial support of Science Foundation Ireland (SFI) and the Department of Agriculture, Food and Marine on behalf of the Government of Ireland (Grant Number [16/RC/3835] - VistaMilk), and the support of US National Science Foundation (NSF) ECCS-2030272, CNS-2212050, and CBET-2316960 grants.

\bibliography{References}
\bibliographystyle{IEEEtran}


\begin{IEEEbiography}[{\includegraphics[width=1in,height=1.25in,clip,keepaspectratio]{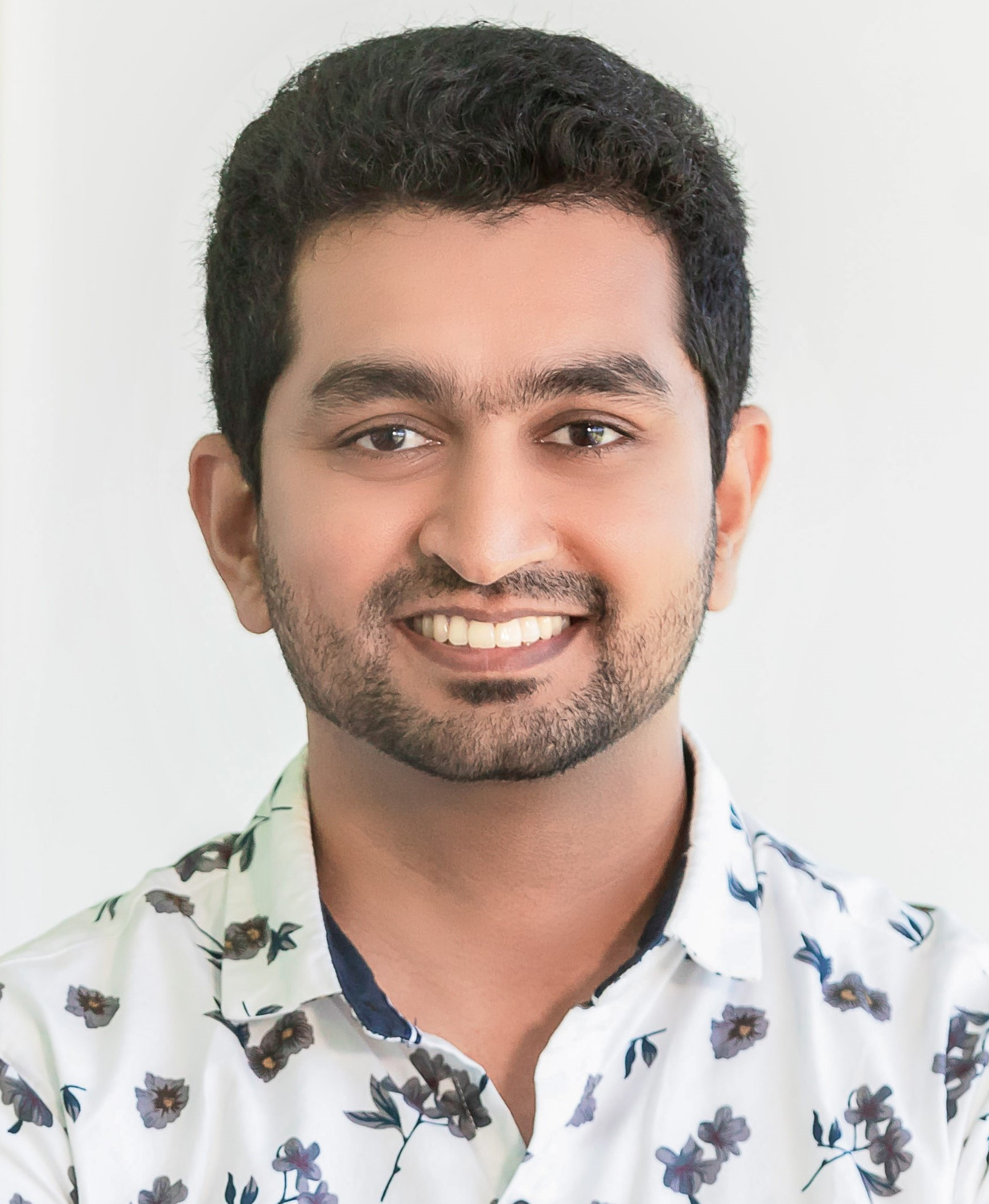}}] {LASANTHA THAKSHILA WEDAGE}
received his B.S. degree in Mathematics from University of Ruhuna, Sri Lanka, in 2016. He is currently pursuing a PhD degree with the Department of Computing and Mathematics at South East Technological University, Ireland. His current research interests include statistics and 5G/6G Wireless communication.
\end{IEEEbiography}
\begin{IEEEbiography}[{\includegraphics[width=1in,height=1.25in,clip,keepaspectratio]{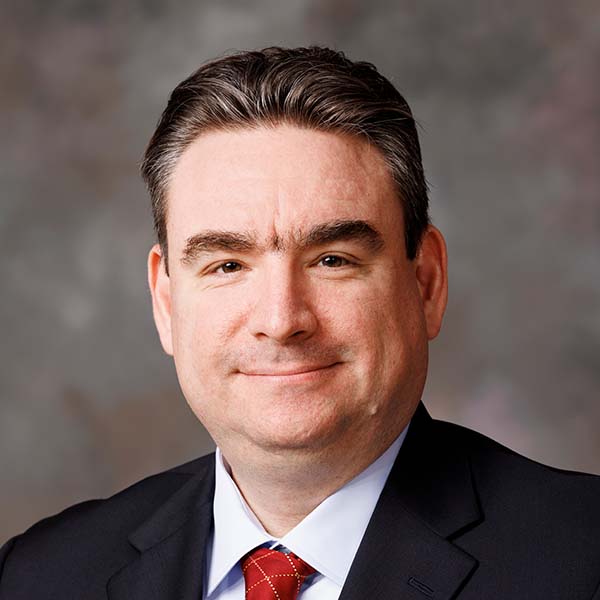}}]{MEHMET CAN VURAN} received his Ph.D. degree in Electrical and Computer Engineering from Georgia Institute of Technology. He is currently the Dale M. Jensen Professor with the School of Computing, University of Nebraska-Lincoln. He is a Daugherty Water for Food Institute Fellow and a National Strategic Research Institute Fellow. His research interests include wireless underground, mmWave, and THz communications in challenging environments, agricultural Internet of Things, dynamic spectrum access, and cyber-physical networking.
\end{IEEEbiography}
\begin{IEEEbiography}[{\includegraphics[width=1in,height=1.25in,clip,keepaspectratio]{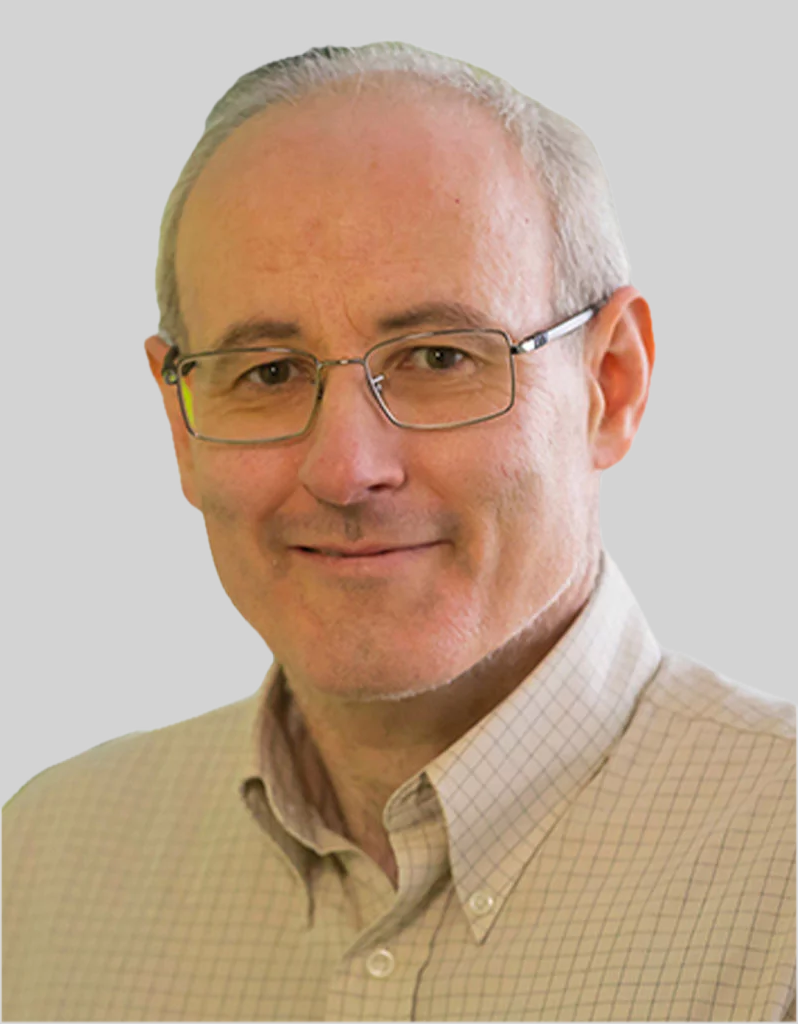}}]{BERNARD BUTLER} received a PhD from WIT, Ireland. He was a Senior Research Scientist in the U.K.’s 
NPL, focusing on the mathematics of measurement and sensing. He lectures at SETU and is a CONNECT Funded Investigator and VistaMilk Academic Collaborator. His research interests include cybersecurity, distributed computing and sensing, applied to future networks, agriculture, transport and health.
\end{IEEEbiography}
\begin{IEEEbiography}[{\includegraphics[width=1in,height=1.25in,clip,keepaspectratio]{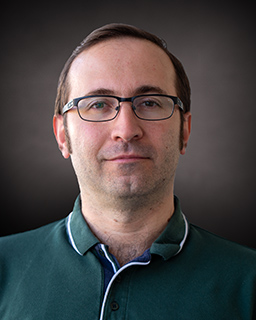}}]{CHRISTOS ARGYROPOULOS} received his Ph.D. degree in Electronic Engineering from Queen Mary, University of London. Currently, he is an Associate Professor with the Department of Electrical Engineering at Pennsylvania State University, USA. He is a senior member of IEEE and an elected Fellow of Optica.
\end{IEEEbiography}
\begin{IEEEbiography}[{\includegraphics[width=1in,height=1.25in,clip,keepaspectratio]{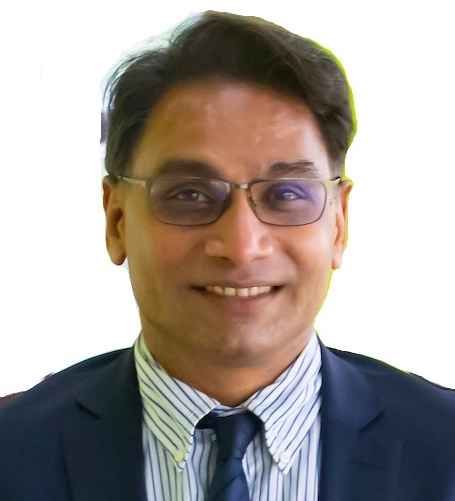}}]{SASITHARAN BALASUBRAMANIAM} received the Bachelor’s degree in engineering and Ph.D. degree from University of Queensland, Brisbane, Australia, in 1998 and 2005, respectively, and the Masters of engineering science from Queensland University of Technology, Brisbane, in 1999. He is currently an Associate Professor with the School of Computing, University of Nebraska-Lincoln, Lincoln, NE, USA. His research interests include molecular/nano communications, Internet of Bio-Nano Things, and 5G/6G networks. 
\end{IEEEbiography}


\end{document}